# X-ray photoelectron emission microscopy in combination with x-ray magnetic circular dichroism investigation of size effects on field-induced Néel-cap reversal

F. Cheynis,[1,2] N. Rougemaille,[1,3] R. Belkhou,[3,4] J.-C. Toussaint,[1,2] and O. Fruchart[1,a]

[1] *Institut NÉEL, CNRS-UJF, BP 166, F-38042 Grenoble Cedex 9, France*
[2] *INPG, 46 Avenue Félix Viallet, F-38031 Grenoble Cedex 1, France*
[3] *ELETTRA, Sincrotrone Trieste, I-34012 Basovizza, Trieste, Italy*
[4] *Synchrotron SOLEIL, L'Orme des Merisiers Saint-Aubin, BP 48, F-91192 Gif-sur-Yvette Cedex, France*

[a] Electronic mail: olivier.fruchart@grenoble.cnrs.fr

**Abstract**

X-ray photoelectron emission microscopy in combination with x-ray magnetic circular dichroism is used to investigate the influence of an applied magnetic field on *Néel caps* (i.e., surface terminations of asymmetric Bloch walls). Self-assembled micron-sized Fe(110) dots displaying a moderate distribution of size and aspect ratios serve as model objects. Investigations of remanent states after application of an applied field along the direction of Néel-cap magnetization give clear evidence for the magnetization reversal of the Néel caps around 120 mT, with a ±20 mT dispersion. No clear correlation could be found between the value of the reversal field and geometrical features of the dots.

Magnetization reversal using an external magnetic field as a driving parameter is a major topic in magnetism. Most studies focus on the reversal of magnetic domains by rotation and/or domain wall propagation. Few reports consider domain walls as object whose inner magnetic structure can be manipulated by a magnetic field. In thin films of soft magnetic material, internal wall structures have been extensively studied by Lorentz microscopy[1,2] and Kerr effect microscopy.[3] However, these systems are easily saturated and a minute in-plane tilt in the magnetic field orientation makes the domain walls propagate. Magnetic nanostructures are systems of growing interest as they constrain geometrically magnetic objects which can be in turn manipulated. Recent reports concern submicron magnetic disks, in which magnetic vortices, i.e., essentially one-dimensional objects, are stabilized. Such vortices have been reversed using a strong out-of-plane magnetic field[4,5] or an in-plane magnetic field of smaller amplitude, either pulsed[6] or oscillating.[7]

Here, we consider an essentially two-dimensional magnetic object: an *asymmetric Bloch wall* of finite length.[3,8,9] The wall was studied in micron-sized elongated self-assembled Fe(110) dots, where a so-called *Landau state* occurs.[10] We evidence that the direction of magnetization in the Néel caps can be reversed by applying a magnetic field around 120 mT along the magnetization of the caps. This has been observed at remanence using x-ray photoelectron emission microscopy in combination with x-ray magnetic circular dichroism (XMCD-PEEM) and reproduced to a good agreement by numerical simulation. In this paper, we put some emphasis on the reversal field distribution, ±20 mT. No clear correlation could be evidenced in relation to the size or geometry of the dots.

The investigation has been performed on self-assembled Fe(110) dots grown by pulsed laser deposition under UHV.[11] These single-crystalline submicron-sized structures are obtained by epitaxial growth on a Mo(110) or W(110) buffer layer deposited on a (11-20) sapphire substrate at typically 500 °C and exhibit atomically flat facets [Fig. 1(a)]. Previous studies[12] have demonstrated both numerically and experimentally that as-grown Fe(110) dots with heights ≥100 nm exhibited the *Landau* state.

In this configuration, the magnetization vector lies essentially in the plane of the sample and displays two main domains along the dot length, antiparallel with each other and thus separated by a 180° Bloch wall. Inside the wall, the magnetization is perpendicularly oriented [Fig. 1(b)] except in the vicinity of both surfaces, where it is terminated by regions with in-plane transverse (i.e., along the width of the nanostructure) magnetization, called in the literature *Néel caps* [Fig. 1(c)]. The two Néel caps are antiparallel [Figs. 1(c) and 1(d)] and this wall is called an *asymmetric Bloch wall*.[8]

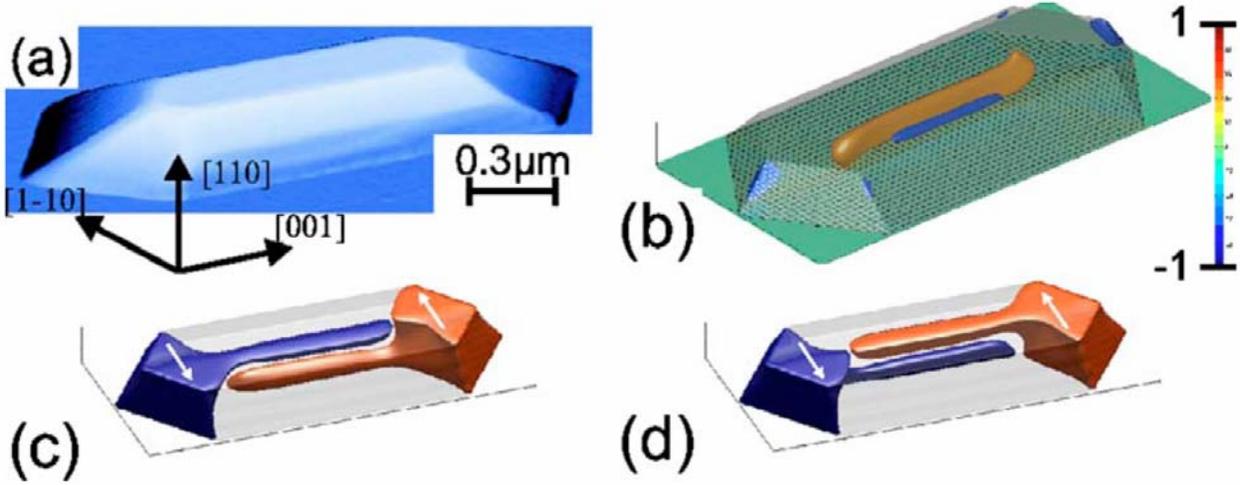

FIG. 1. (Color online) (a) Typical AFM image of an elongated Fe(110) dot. [(b) and (c)] Micromagnetic simulations of the Landau state in a facetted Fe(110) dot ($1000 \times 500 \times 100$ nm$^3$). In (b), regions with a strong perpendicular component of the magnetization ($|m_z| \geq 0.5$) are displayed, showing the Bloch wall. In (c) and (d), only the transverse magnetization component is represented, highlighting the antiparallel Néel caps.

We used the French X-PEEM instrument installed at the nanospectroscopy beamline of the synchrotron Elettra (Trieste, Italy) in combination with x-ray magnetic circular dichroism (XMCD-PEEM). Mirror electron microscopy (MEM) has been used to determine the topography of dots. XMCD-PEEM technique enables, with high spatial resolution ($\approx 20$ nm), to image the surface magnetization distribution along the incident x-ray beam. Using this technique, top surface Néel caps of similar Fe(110) dots have been evidenced in a previous study.[12] However, as the PEEM technique is based on the collection of electrons with low energy, only very moderate in-plane magnetic field could be applied. To overcome this limitation, a statistical approach has been adopted. We have applied a transverse field *ex situ*, introduced the sample under UHV, and imaged it at remanence. This manipulation has been repeated for each value of the applied field. Typically, 30–40 dots have been imaged for each field value to determine with a sufficient statistics the ratio of each top Néel-cap population (black or white in our images) to the total Néel-cap population.

Starting from an as-grown assembly of Fe(110) dots, the ratio of the white Néel caps to the total number of dots imaged at remanence increases with the applied field until it reaches ~95% for an applied field of 150 mT [Fig. 2(c)]. The mean reversal field is 120±20 mT. We have also checked that applying a magnetic field of −150 mT yielded a population inversion of the Néel caps (ratio of black Néel caps of ≥95%). We have thus evidenced that an internal component of a domain wall can be controlled by a magnetic field.

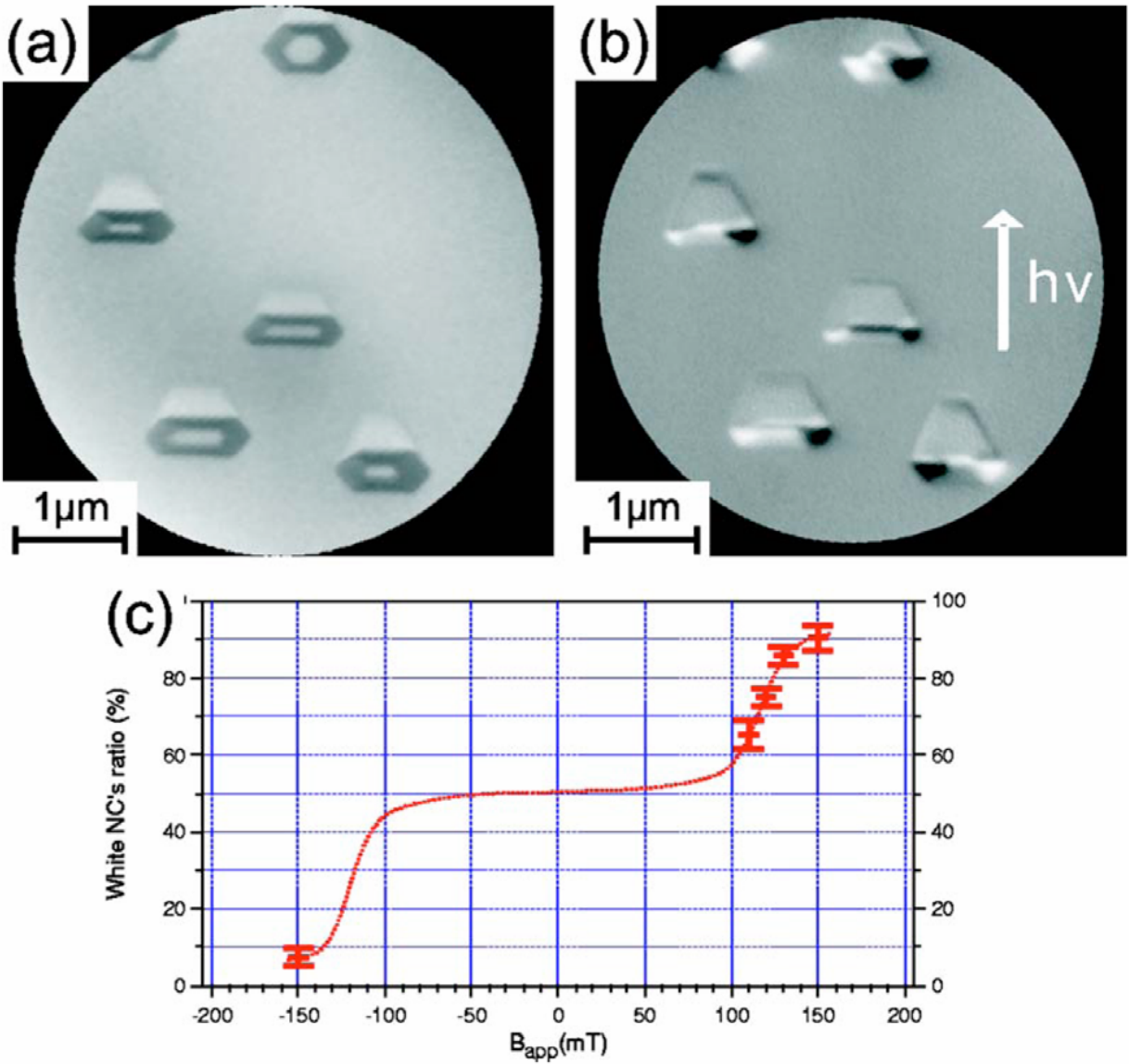

FIG. 2. (Color online) [(a) and (b)] Typical views of Fe(110) dots: (a) MEM images (topography) and (b) XMCD-PEEM images (magnetism). In (b), the white arrow indicates the x-ray beam direction. (c) Evolution of the white Néel cap (NC) ratio as a function of the transverse magnetic field.

As XMCD-PEEM gives access to the top surface component of the magnetization only, three-dimensional micromagnetic simulations are required to understand the details of the reversal. This is reported in Ref. 13. Here, we will repeat only the information needed to follow the main topic of the article, which is the investigation of the distribution of reversal field. We used a custom-made micromagnetic code based on the temporal integration of the Landau-Lifschitz-Gilbert equation, i.e., in a continuous medium approximation. These simulations are performed with the finite difference method. In this framework, the sample is discretized into prisms of regular size. The micromagnetic simulations predict that at remanence, the Néel caps are always antiparallel to each other, which means that during the reversal process, both caps reversed their magnetization. This is accomplished by the reversal of the cap with magnetization antiparallel to the field for rising field and by the reversal of systematically the top cap for decreasing field. The cap observed by XMCD-PEEM therefore has its magnetization opposite to the applied field. The experimental and numerical reversal fields are in good agreement.

Let us now investigate whether the reversal field depends on geometrical parameters. Several arguments could be put forward. First, we could expect that increasing the nanostructure height would result in decreasing the dipolar coupling between the two Néel caps which would in turn decrease the field value needed to reverse the Néel caps. Second, the reversal field may also depend on the vertical aspect ratio (height/width). Indeed, if the vertical aspect ratio increases, then the transverse demagnetizing coefficient also increases, so that the magnetization in the two longitudinal domains rotates less under a given transverse field, and finally the field value at which the two caps become parallel is increased. The reason is, following the variational model first proposed by Labonte,[8] that the cost in exchange energy of the Néel cap antiparallel to the field, i.e., for which a rotation larger than 180° is required, is lowered as the angle between the longitudinal domains is also reduced. Thus, an increase of the reversal field of the cap would be expected.

For each applied field, we have studied the total Néel cap population (black and white) and the population of white Néel caps as a function of height, lateral aspect ratio (length/width), and vertical aspect ratio (height/width). The length and width of both top and bottom surfaces have been determined from MEM images (Fig. 3). Each distance has been measured using intensity profiles and defined at half-height. For dots displaying asymmetry, maximum projected lengths of both surfaces have been considered. As the angle between the (110) base plane and the

(010) planes that define the longest sides of a dot equals 45°, the height of a dot is computed as the half-difference between the bottom and top surface widths. Typical relative uncertainty on distances is 7 nm.

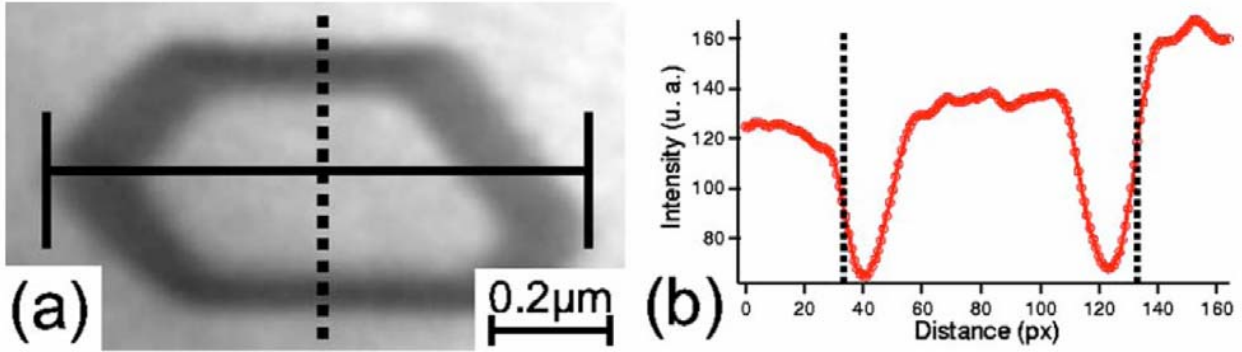

FIG. 3. (Color online) Illustration of distance measurements. (a) MEM image of an Fe(110) dot displaying asymmetry. The bottom surface length is defined between the two vertical black lines. (b) Intensity profile corresponding to the black dotted line in (a). The bottom surface width is defined at half-height (black vertical dotted lines) of the intensity profile.

Figure 4 illustrates the data that we have obtained as a function of the height (left) and the vertical aspect ratio (right) of nanostructures (similar results have been obtained for the lateral aspect ratio, not shown here). In the case of a dependence of the reversal field on these parameters, we would have expected, for increasing field values, the onset of an asymmetry in Figs. 4(a)–4(c) and different saturation fields (i.e., field value at which all imaged dots are white) in Fig. 4(d). However, due to statistical uncertainties, the different curves overlap. To determine these uncertainties, we have supposed that all white Néel-cap histogram bin values are random variables with a standard deviation that equals $\sqrt{N_i^{tot}}$, where $N_i^{tot}$ is the total number of imaged dots of a given histogram bin $i$. Uncertainty on white Néel-cap ratios is then $1/\sqrt{N_i^{tot}}$.

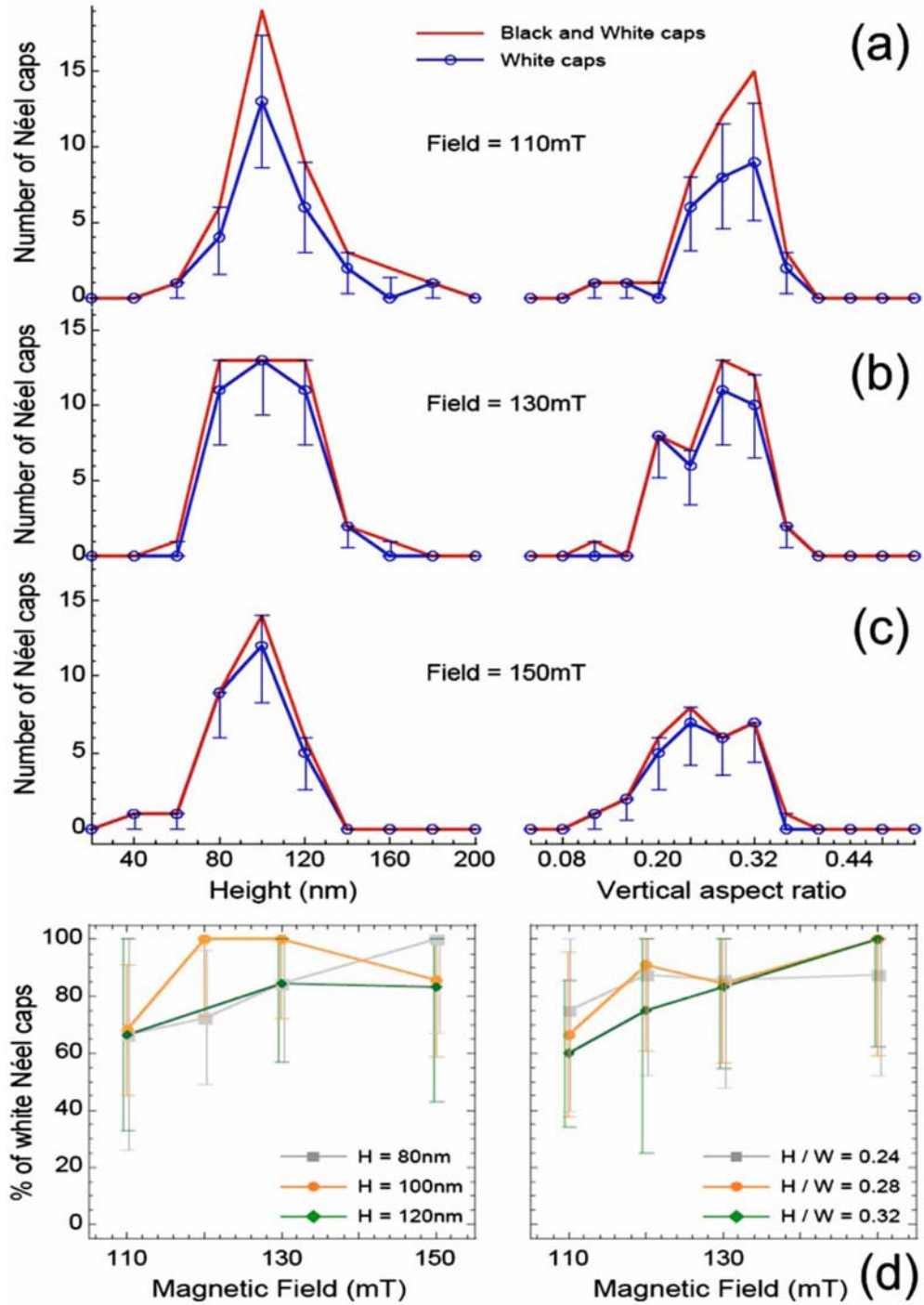

FIG. 4. (Color online) Example of data postprocessing. [(a)–(c)] Height (left) and vertical aspect ratio (right) histograms of the total Néel-cap population (solid line) and the white Néel-cap population (open circles) for different applied fields. (d) Computed from (a) to (c): for different heights (left) and different vertical aspect ratios (right), ratios of the white Néel-cap population to the black and white Néel-cap population as a function of the applied field.

These overlaps entail that, in the present study, no clear correlation between reversal fields and geometrical parameters can be evidenced. This means that either the reversal field is independent of geometrical considerations or our statistics was not sufficient to evidence this effect. In the former case, effects connected with the three-dimensional nature of the magnetic configuration, such as the magnetization reversal at edges, may be involved.

In conclusion, using self-assembled Fe(110) dots as model systems, we have shown with XMCD-PEEM microscopy that the orientation of internal features of an asymmetric Bloch wall, known as Néel caps, can be reversed by applying a magnetic field transverse to the dot. A mean reversal field of 120 mT has been evidenced, in good agreement with numerical simulations. The reversal field distribution of ±20 mT could not be correlated to any geometrical parameter. Its origin is still an open question.

[1] S. Tsukahara and H. Kawakatsu, J. Phys. Soc. Jpn. **32**, 1493 (1972).

[2] L. Zepper and A. Hubert, J. Magn. Magn. Mater. **2**, 18 (1976).

[3] A. Hubert and R. Schäfer, *Magnetic Domains: The Analysis of Magnetic Microstructures* (Springer, New York, 1998).

[4] T. Okuno, K. Shigeto, T. Ono, K. Mibu, and T. Shinjo, J. Magn. Magn. Mater. **240**, 1 (2002).

[5] A. Thiaville, J. M. García, R. Dittrich, J. Miltat, and T. Schrefl, Phys. Rev. B **67**, 094410 (2003).

[6] R. Hertel, S. Gliga, C. Schneider, and M. Fähnle, Phys. Rev. Lett. **98**, 117201 (2007).

[7] B. Van Waeyenberg, A. Puzic, H. Stoll, K. W. Chou, T. Tyliszczak, R. Hertel, M. Fähnle, H. Brückl, K. Rott, G. Reiss, I. Neudecker, D. Weiss, C. H. Back, and G. Schütz, Nature (London) **444**, 461 (2006).

[8] A. E. Labonte, J. Appl. Phys. **40**, 2450 (1969).

[9] A. Hubert, Phys. Status Solidi **32**, 519 (1969).

[10] P.-O. Jubert, J.-C. Toussaint, and O. Fruchart, Europhys. Lett. **63**, 135 (2003).

[11] O. Fruchart, P.-O. Jubert, M. Eleoui, F. Cheynis, B. Borca, P. David, V. Santonacci, A. Liénard, M. Hasegawa, and C. Meyer, J. Phys.: Condens. Matter **19**, 053001 (2007).

[12] R. Hertel, O. Fruchart, S. Cherifi, P.-O. Jubert, S. Heun, A. Locatelli, and J. Kirschner, Phys. Rev. B **72**, 214409 (2005).

[13] F. Cheynis, O. Fruchart, J.-C. Toussaint, N. Rougemaille, and R. Belkhou (unpublished).